# Brain Modelling as a Service: The Virtual Brain on EBRAINS

Michael Schirner[a,b,c,d,e], Lia Domide[f], Dionysios Perdikis[a,b], Paul Triebkorn[a,b,g], Leon Stefanovski[a,b], Roopa Pai[a,b], Paula Prodan[f], Bogdan Valean[f], Jessica Palmer[a,b], Chloê Langford[a,b], André Blickensdörfer[a,b], Michiel van der Vlag[h], Sandra Diaz-Pier[h], Alexander Peyser[h], Wouter Klijn[h], Dirk Pleiter[i,j], Anne Nahm[i], Oliver Schmid[k], Marmaduke Woodman[g], Lyuba Zehl[l], Jan Fousek[g], Spase Petkoski[g], Lionel Kusch[g], Meysam Hashemi[g], Daniele Marinazzo[m,n], Jean-François Mangin[o], Agnes Flöel[p,q], Simisola Akintoye[r], Bernd Carsten Stahl[s], Michael Cepic[t], Emily Johnson[t], Gustavo Deco[u,v,w,x], Anthony R. McIntosh[y], Claus C. Hilgetag[z, α], Marc Morgan[k], Bernd Schuller[i], Alex Upton[β], Colin McMurtrie[β], Timo Dickscheid[l], Jan G. Bjaalie[χ], Katrin Amunts[l,δ], Jochen Mersmann[ε], Viktor Jirsa[g], Petra Ritter[a,b,c,d,e,*]

[a]Berlin Institute of Health at Charité – Universitätsmedizin Berlin, Charitéplatz 1, 10117 Berlin, Germany
[b]Charité – Universitätsmedizin Berlin, corporate member of Freie Universität Berlin and Humboldt Universität zu Berlin, Department of Neurology with Experimental Neurology, Charitéplatz 1, 10117 Berlin, Germany
[c]Bernstein Focus State Dependencies of Learning & Bernstein Center for Computational Neuroscience, Berlin, Germany
[d]Einstein Center for Neuroscience Berlin, Charitéplatz 1, 10117 Berlin, Germany
[e]Einstein Center Digital Future, Wilhelmstraße 67, 10117 Berlin, Germany
[f]Codemart, Str. Petofi Sandor, Cluj Napoca, Romania
[g]Institut de Neurosciences des Systèmes, Aix Marseille Université, Marseille, France
[h]SimLab Neuroscience, Jülich Supercomputing Centre (JSC), Institute for Advanced Simulation, JARA, Forschungszentrum Jülich GmbH, Jülich, Germany
[i]Jülich Supercomputer Centre, Forschungszentrum Jülich, Jülich, Germany
[j]Institut für Theoretische Physik, Universität Regensburg, Regensburg, Germany
[k]Swiss Federal Institute of Technology Lausanne (EPFL), Campus Biotech, Geneva, Switzerland
[l]Institute of Neuroscience and Medicine (INM-1), Research Centre Jülich, Jülich, Germany
[m]Department of Data-Analysis, Faculty of Psychology and Educational Sciences, Ghent University, Belgium
[n]San Camillo IRCSS, Venice, Italy
[o]Université Paris-Saclay, CEA, CNRS, Neurospin, Baobab, Gif-sur-Yvette, France
[p]Department of Neurology, University Medicine Greifswald, Greifswald, Germany
[q]German Center for Neurodegenerative Diseases (DZNE) Standort Greifswald, Greifswald, Germany
[r]Leicester De Montfort Law School, De Montfort University, Leicester, United Kingdom
[s]Centre for Computing and Social Responsibility, De Montfort University, Leicester, United Kingdom
[t]Department of Innovation and Digitalisation in Law, Faculty of Law, University of Vienna, Austria
[u]Center for Brain and Cognition, Computational Neuroscience Group, Department of Information and Communication Technologies, Universitat Pompeu Fabra, Barcelona, Spain
[v]Institució Catalana de la Recerca i Estudis Avançats, Barcelona, Spain
[w]Department of Neuropsychology, Max Planck Institute for Human Cognitive and Brain Sciences, Leipzig, Germany
[x]School of Psychological Sciences, Turner Institute for Brain and Mental Health, Monash University, Melbourne, Clayton, Australia
[y]Baycrest Health Sciences, Rotman Research Institute, Toronto, ON, Canada
[z]Institute of Computational Neuroscience, University Medical Center Hamburg-Eppendorf, Hamburg, Germany
[α]Department of Health Sciences, Boston University, 635 Commonwealth Ave., Boston, Massachusetts 02215
[β]CSCS Swiss National Supercomputing Centre, Lugano, Switzerland
[χ]Institute of Basic Medical Sciences, University of Oslo, Norway
[δ]C. and O. Vogt Institute for Brain Research, University Hospital Düsseldorf, Heinrich-Heine University Düsseldorf, Germany
[ε]CodeBox GmbH, Stuttgart, Germany
[*]Corresponding author: Petra Ritter, petra.ritter@charite.de, Charitéplatz 1, 10117 Berlin, Germany

Keywords: brain modelling, cloud, connectome, neuroimaging, network model, high performance computing

**The Virtual Brain (TVB) is now available as open-source cloud ecosystem on EBRAINS, a shared digital research platform for brain science. It offers services for constructing, simulating and analysing brain network models (BNMs) including the TVB network simulator; magnetic resonance imaging (MRI) processing pipelines to extract structural and functional connectomes; multiscale co-simulation of spiking and large-scale networks; a domain specific language for automatic high-performance code generation from user-specified models; simulation-ready BNMs of patients and healthy volunteers; Bayesian inference of epilepsy spread; data and code for mouse brain simulation; and extensive educational material. TVB cloud services facilitate reproducible online collaboration and discovery of data assets, models, and software embedded in scalable and secure workflows, a precondition for research on large cohort data sets, better generalizability and clinical translation.**

Scientific studies are often difficult to replicate and findings often do not generalize in light of additional data (Aarts et al., 2015; Ioannidis, 2005). The data and the computational steps that produced the findings as well as the explicit workflow describing how to generate the results were identified as the minimal components for independent regeneration of computational results (Stodden et al., 2016). EBRAINS (European Brain Research INfrastructureS) is an open brain research platform that makes data, tools and results accessible to everyone within an environment that promotes reproducible work. It offers cloud services for collaborative online research, databases with annotated and curated data of many modalities, atlases of human and rodent brains, data processing workflows, supercomputing resources, neuromorphic systems, and virtual robots. EBRAINS was developed by the Human Brain Project (HBP), a research initiative funded by the European Commission with the mission to decode the human brain (Amunts et al., 2019, 2016). TVB cloud services (Tables 1, 2) were developed by the HBP subproject "The Virtual Brain" in collaboration with the two HBP partnering projects TVB-Cloud (virtualbraincloud-2020.eu) and TVB-CD (bit.ly/3ogLYtb). To provide supercomputing resources, the HBP offers as part of the Interactive Computing E-Infrastructure project access to compute and storage resources of the Fenix infrastructure (fenix-ri.eu), a network of five European supercomputing centres.

TVB services use EBRAINS core services for deployment (Figure 1). The 'Collaboratory' (Supplementary Note: The EBRAINS Collaboratory) hosts light-weight workspaces, called 'collabs', where research teams can exchange data and work together on Office or Wiki documents, secured with access control. 'Lab' provides sandboxed JupyterLab instances for developing applications and running code. Lab notebooks provide a powerful interface to EBRAINS services, for data transfer as well as configuring and executing jobs on supercomputers. Data can be found and accessed via the 'KnowledgeGraph' (KG), which provides a graphical user interface (GUI) and Application Programming Interface (API) for searching, inserting, editing, querying, aggregating, filtering and visual exploration of the data base. KG uses controlled vocabularies and ontologies, mapped with existing neuroimaging and brain simulation ontologies (Supplementary Note: Metadata annotations). Professional curation, persistent DOIs, licensing, versioning, data sharing agreements and protected workflows enable secure, interoperable and reusable sharing of data and services. Containerized workflows and backend supercomputing resources enable reproducible research that scales to the demands of the project. A RESTful API is used for connecting different cloud components, authentication, data transfer and supercomputer job scheduling. Atlas Services provide common spatial reference spaces including a multilevel atlas of the human brain as well as the Waxholm Space rat brain atlas (Osen et al., 2019; Papp et al., 2014). The Multilevel Human Brain Atlas uses the Julich-Brain probabilistic cytoarchitectonic maps (Amunts et al., 2020) to link with template spaces such as BigBrain (Amunts et al., 2013) at the micrometer scale and MNI (Das et al., 2016) at millimeter scale, and combines them with imaging-based maps of function (Evans et al., 2012) and connectivity (Guevara et al., 2017) to link a growing set of multimodal feature descriptions of the human brain, in order to capture brain organization in its different facets.

Results

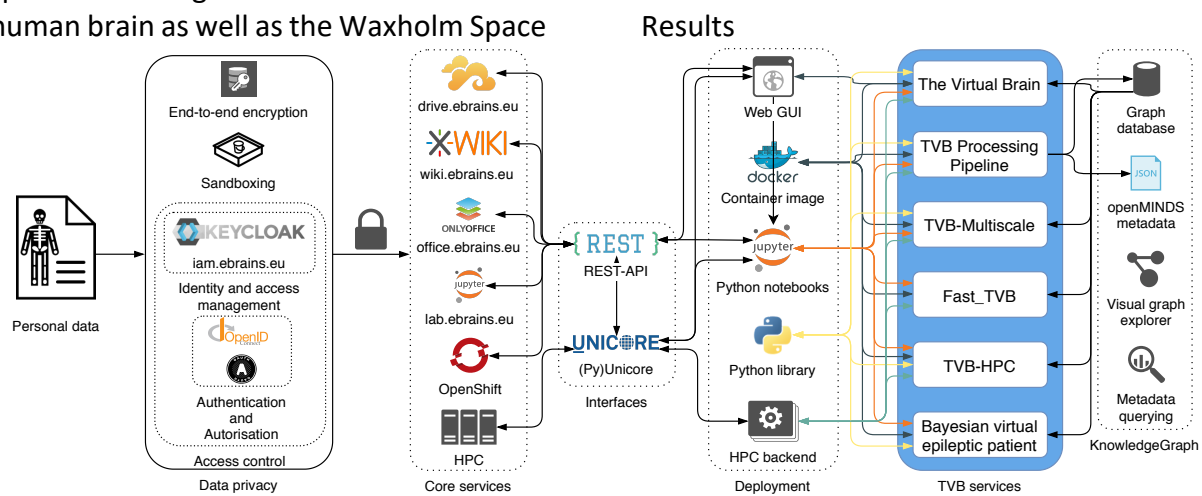

Figure 1. TVB on EBRAINS cloud ecosystem. Brain simulation and neuroimaging require personal data applicable to data protection regulation. Encryption, sandboxing and access control are used to protect personal data. EBRAINS provides core services: 'Drive' for hosting and sharing files; 'Wiki' and 'Office' to create workspaces and documents for collaborative research; 'Lab' for running live code in sandboxed JupyterLab instances; 'OpenShift' for orchestrating services and resource management; 'HPC' are supercomputers for resource-intensive computations. Services interact via RESTful APIs and use UNICORE for communication with supercomputers. Services are deployed in the form of Web GUIs, container images, Python notebooks, Python libraries and high-performance machine codes. Input and output data can be loaded from and stored into the EBRAINS Knowledge Graph using openMINDS-compliant metadata annotations to enable efficient sharing and re-use. The connectors show interactions between different components (colours group connectors for different deployments).

| Service | Function | URLs |
|---|---|---|
| The Virtual Brain | Brain network simulation | *Web-App* |
| | | thevirtualbrain.apps.hbp.eu |
| | | *Collab* |
| | | wiki.ebrains.eu/bin/view/Collabs/the-virtual-brain |
| | | *End-to-end use case*<br>  wiki.ebrains.eu/bin/view/Collabs/user-story-tvb |
| | | *Source code*<br>  github.com/the-virtual-brain/tvb-root |
| | | *Python libraries*<br>  tvb-library<br>  tvb-framework |
| | | *Container image*<br>  hub.docker.com/r/thevirtualbrain/tvb-run |
| TVB Image Processing Pipeline | Connectome analysis | *Web-App*<br>  tvb-pipeline.apps.hbp.eu |
| | | *Collab*<br>  wiki.ebrains.eu/bin/view/Collabs/tvb-pipeline |
| | | *Source codes*<br>  github.com/BrainModes/tvb-pipeline-sc<br>  github.com/BrainModes/fmriprep<br>  github.com/BrainModes/tvb-pipeline-converter |
| | | *Container images*<br>  hub.docker.com/r/thevirtualbrain/tvb-pipeline-sc<br>  hub.docker.com/r/thevirtualbrain/tvb-pipeline-fmriprep<br>  hub.docker.com/r/thevirtualbrain/tvb-pipeline-converter |
| Multiscale co-simulation | Two toolboxes for concurrent simulation of large-scale and spiking networks | *Web-App (TVB-Multiscale)*<br>  tvb-nest.apps.hbp.eu |
| | | *Collab (TVB-Multiscale)*<br>  wiki.ebrains.eu/bin/view/Collabs/the-virtual-brain-multiscale |
| | | *Collab (Parallel CoSimulation)*<br>  wiki.ebrains.eu/bin/view/Collabs/co-simulation-tvb-and-nest-high-computer |
| | | *Source code (TVB-Multiscale)*<br>  github.com/the-virtual-brain/tvb-multiscale |
| | | *Source code (Parallel CoSimulation)*<br>  github.com/multiscale-cosim/TVB-NEST |
| | | *Container image (TVB-Multiscale)*<br>  hub.docker.com/r/thevirtualbrain/tvb-nest |
| TVB-HPC | Automatic code generation | *Collab*<br>  wiki.ebrains.eu/bin/view/Collabs/rateml-tvb/ |
| | | *Source code*<br>  github.com/the-virtual-brain/tvb-root |
| Fast_TVB | Parallelized ReducedWong Wang | *Collab*<br>  wiki.ebrains.eu/bin/view/Collabs/fast-tvb |
| | | *Source code*<br>  github.com/BrainModes/fast_tvb |
| | | *Container image*<br>  hub.docker.com/r/thevirtualbrain/fast_tvb |
| Bayesian Virtual Epileptic Patient | Epilepsy modelling | *Collab*<br>  wiki.ebrains.eu/bin/view/Collabs/bayesian-virtual-epileptic-patient |
| | | *Source code*<br>  github.com/ins-amu/BVEP |
| TVB Mouse Brains | Mouse brain simulation | *Collabs*<br>  wiki.ebrains.eu/bin/view/Collabs/tvb-mouse-brains<br>  wiki.ebrains.eu/bin/view/Collabs/mouse-stroke-brain-network-model/ |
| TVB-ready dataset | SC, FC, and fMRI from tumour patients and controls | *DOI*<br>  10.25493/1ECN-6SM<br>*URL*<br>  kg.ebrains.eu/search/instances/Dataset/a696ccc7-e742-4301-8b43-d6814f3e5a44 |
| openMINDS metadata for TVB-ready data | Metadata in JSON-LD format | *Collab*    wiki.ebrains.eu/bin/view/Collabs/openminds-metadata-for-tvb-ready-data |
| | | *openMINDS schema*    github.com/HumanBrainProject/openMINDS |
| TVB atlas adapter | Brain atlas | *Collab (development version)*    wiki.ebrains.eu/bin/view/Collabs/sga3-d1-1-showcase-1 |
| | | *Visualizer*    brainsimulation.org/atlasweb_multiscale |
| INCF TVB training space | Education and training | *URL*    training.incf.org/collection/virtual-brain-simulation-platform |

Table 1. TVB cloud services and URLs leading to their main entry points.

| Cloud service | Publications |
|---|---|
| The Virtual Brain | (Ritter et al., 2013; Sanz-Leon et al., 2015, 2013) |
| TVB Image Processing Pipeline | (Proix et al., 2016; Schirner et al., 2015) |
| Fast_TVB | (Costa-Klein et al., 2020; Schirner et al., 2018; Shen et al., 2019; Zimmermann et al., 2018) |
| Bayesian Virtual Epileptic Patient | (Hashemi et al., 2020; Jirsa et al., 2017) |
| TVB Mouse Brain | (Melozzi et al., 2019, 2017) |
| TVB ready datasets | (Aerts et al., 2020, 2018) |
| INCF TVB training space | (Matzke et al., 2015) |

Table 2. Exemplary publications using software, workflow or data sets underlying different TVB cloud services.

The Virtual Brain

TVB (thevirtualbrain.org) is an open-source BNM simulator that combines experimental data with neuron theory for brain research (Supplementary Note: Brain simulation with TVB) (Ritter et al., 2013; Sanz-Leon et al., 2013). A BNM is a computer model for simulating brain activity based on systems of differential equations that are coupled by a reconstruction of a brain's structural connectome (SC), the white-matter axon fiber bundle network that interconnects brain areas (Breakspear, 2017). Neural populations are simulated by neural mass models (NMM), i.e., mean field theories on coupled neurons. SCs can be reconstructed from diffusion-weighted MRI (dwMRI) data using the TVB Image Processing Pipeline or found with KG.

BNMs simulate neural activity like membrane voltage fluctuations, synaptic current flow, or spike-firing, which is used to predict signals like functional MRI (fMRI) or electroencephalograms (EEG). Predicted signals can be compared with empirical measurements to evaluate and optimize the model or to analyse the underlying model mechanisms. TVB can be directly used from a web GUI (Table 1), without the need to install further software or to have a specific operating system, computing environment or hardware. In addition, compiled standalone versions for Linux, Windows and Mac OS as well as execution-ready container images are freely available for download. TVB can also be used as a Python library for programming--locally, as well as in the EBRAINS Lab (Figure 1). TVB usage is introduced through Jupyter notebooks, explanatory videos and technical documentation. The main documentation is hosted at docs.thevirtualbrain.org.

TVB Image Processing Pipeline

The TVB Image Processing Pipeline takes anatomical, functional and diffusion MRI as input and generates SCs, region-average fMRI time series, functional connectivity (FC), brain surface triangulations, projection matrices for predicting EEG, and brain parcellations as output. The outputs can be used for analysis or for direct upload to TVB. Processing steps are performed by containerized workflows, orchestrated by Python scripts, for flexible adaptation. The pipeline combines three BIDS Apps ("Brain Imaging Data Structure", (Gorgolewski et al., 2016)) that can be run via command-line interface. The container codes, respectively images, received version numbers and are hosted at GitHub, respectively Docker Hub, for change tracking (Table 1). Jupyter notebooks explain how the TVB Image Processing Pipeline can be executed on a supercomputer from a web browser using access control, encryption and sandboxing for protection of personal data.

Multiscale Co-Simulation

Multiscale Co-Simulation are two Python toolboxes for simulating brain networks where large-scale NMMs interact with models of individual neurons or neuron networks (Table 1). TVB is used to simulate the large-scale NMMs, while NEST (Gewaltig and Diesmann, 2007) is used to simulate the small-scale neurons. The toolboxes provide two interfaces to couple the two simulators, using the programmatic Python interface of TVB and PyNEST (Eppler et al., 2009), a Python wrapper for NEST. Populations interact with neurons by coupling NMM state variables with single neuron state variables or parameters. For example, a TVB state variable that simulates ongoing population firing can be used to inject spikes into a NEST spiking network, e.g., by sampling spike times from a probability distribution in dependence of the instantaneous firing rate of the NMM. Vice versa, the mean activity of a NEST neuron network may be used to inform ongoing inputs to a TVB NMM. Coupling may be unidirectional, e.g., to study effects of large-scale inputs on small-scale spiking-network activity, or bidirectional, to study how both scales mutually interact. The TVB-multiscale project is under ongoing development currently focussing on postulating and validating coupling scenarios between the scales, optimizing the user interfaces as well as optimizing performance via parallelization and better interprocess communication. TVB-multiscale can be downloaded as standalone container image or used on EBRAINS from Jupyter notebooks (Table 1).

High-Performance implementations of TVB

Brain modelling often requires the exploration of high-dimensional parameter spaces and thus simulation software need to be efficient and able to run on parallel architectures. Numba, a compiler that translates a subset of Python and NumPy into high-performance machine code, is used for faster execution of TVB. However, the central integration loop is implemented in Python for modularity and generality and therefore constitutes a bottleneck for simulation speed, which is why two dedicated high-performance implementations were created.

TVB-HPC automatically produces high-performance codes for CPUs and GPUs using the domain-specific language RateML for model specification in order to simplify the process of implementing optimized BNM simulation codes. RateML is based on the domain-independent language 'LEMS' (Vella et al., 2014), which allows for the declarative description of model components in a concise XML representation. TVB-HPC is part of the main TVB Python toolbox (Table 1).

Fast_TVB is a specialized high-performance implementation of the ReducedWongWang model (Deco et al., 2014), written in C, that makes use of "Single Instruction, Multiple Data" operations, explicit memory management and C pointers to efficiently use CPU resources. Fast_TVB's high efficiency (Supplementary Figure 6) enables to perform dense parameter space explorations or to simulate high-dimensional models with millions of nodes. The code uses multithreading to simultaneously perform the processing of one BNM on multiple processors. Fast_TVB is deployed as a standalone container image (Table 1).

Bayesian Virtual Epileptic Patient

The Bayesian Virtual Epileptic Patient (BVEP) uses Bayesian inference to compute posterior probability distributions for parameters of TVB's Epileptor NMM in order to study the spread of epileptic seizures (Jirsa et al., 2017, 2014). The approach applies Bayes' theorem on prior distributions obtained from empirical data (e.g., a patient's SC, or lesions detected in MRI) and model simulations, taking into account the likelihood for these observations. Estimating the excitability parameter of an Epileptor-BNM for every brain region yields a map of epileptogenicity to guide clinical decision-making. The excitability parameter controls whether an Epileptor NMM shows epileptogenic behavior and the estimated values are used to classify brain regions into three categories: epileptogenic zones (EZ), which can autonomously trigger seizures; propagation zones (PZ), which do not trigger seizures autonomously but may be recruited during the seizure evolution; and healthy zones (HZ), where no seizures occur. Priors for the excitability parameter can express clinical hypotheses or empirical observations (e.g., in a seizure region the HZ range can be excluded from the prior). Simulation results are compared with empirical measurements (e.g., EEG from implanted electrodes) and model selection with cross-validation metrics are computed to evaluate clinical hypotheses and to assess the model's ability to predict new data. The workflow was successfully used to infer the spatial maps of epileptogenicity from ground-truth synthetic data for symptomatic and asymptomatic seizures (Hashemi et al., 2020). The currently running EPINOV clinical trial (epinov.com) investigates informing clinical decisions with virtual patient studies to improve surgery outcome. To run BVEP, users may follow the instructions in the collab or the source code repository (Table 1).

TVB Mouse Brains

The Virtual Mouse Brain (TVMB) extends TVB with tractography-based as well as tracer-based mouse SC (Melozzi et al., 2017). Tracer-based SC was exported from the Allen Mouse Brain Connectivity Atlas (Oh et al., 2014) using the Allen Connectivity Builder. Two use cases of TVMB are demonstrated on EBRAINS (Table 1). The first use case performs a bifurcation analysis to show the existence of multistability in mouse BNMs and compares FC simulated using tracer-based SC versus FC using dwMRI-based SC (Melozzi et al., 2019). The second use case ("Stroke Mouse Brain") constructs custom mouse SC at different resolutions (the maximum resolution of 50 μm yields 540 brain regions), fits the working point of a mouse BNM and explores ways to simulate stroke and rehabilitation in mice (Allegra Mascaro et al., 2020).

### TVB-ready data

The EBRAINS KG provides users with TVB-ready reference data sets in BIDS format from tumor patients and matched control participants. The data set contains region-average fMRI time series, FC, and SC from 31 brain tumor patients before and after surgery, and 11 healthy controls (Aerts et al., 2019). Planning for tumor surgery involves delineating eloquent tissue to spare by analyzing neuroimaging data like fMRI and dwMRI data. Instead of analyzing these modalities independently, BNMs provide a novel way to combine their information.

Optimized parameters in presurgical models differentiated between regions directly affected by a tumor, regions distant from a tumor and regions in a healthy brain, which may help to better delineate eloquent tissue (Aerts et al., 2018). Furthermore, it was found that "virtual neurosurgeries", where the patients' actual surgeries were simulated, improved the fit with postsurgical brain dynamics, which may allow presurgical exploration of different surgical strategies (Aerts et al., 2020).

### TVB atlas and data adapters

Adapters connect TVB with data and software (Table 1; Supplementary Note: TVB atlas and data adapters). Among others, TVB has interfaces with MATLAB, the Brain Connectivity Toolbox, and the Allen Brain Atlas (docs.thevirtualbrain.org). Adapters are under development that connect TVB with the Human Brain Atlas, Knowledge Graph and Human Intracerebral EEG Platform to inform BNM parameterization and to compare simulation results with empirical data (Table 1). The Human Brain Atlas characterizes brain regions with a growing set of multimodal data features, including transmitter receptor densities (Palomero-Gallagher and Zilles, 2019), cell distributions, and physiological recordings, based on the Julich-Brain cytoarchitectonic maps (Amunts et al., 2020). Aligned with standard brain templates, the Human Atlas can be registered with individual brains to export multimodal microstructural "fingerprints" that can be used to set the parameters of BNM nodes. For example, density measurements for 16 receptors will be provided for each brain region, and high-resolution tractography maps (full brain, post-mortem 200 μm isotropic diffusion MRI and 60 μm isotropic 3D Polarised Light Imaging) will increase the reliability of SC (Axer et al., 2016; Beaujoin et al., 2018). As data adapter, it is planned to link intracranial electrophysiology recordings with the respective Julich-Brain regions to set parameters based on direct measurements of effective connectivity and transmission delays from stimulation experiments (Trebaul et al., 2018). A viewer was implemented to visualize different atlas maps on the cortical surface (Table 1).

### INCF training space

The INCF (International Neuroinformatics Coordination Facility) training space holds a dedicated collection for TVB with didactic use cases, video tutorials, notebooks and example data sets (Supplementary Note: INCF training space). INCF's TVB EduPack module (Table 1) gives a thorough introduction into work with TVB in general and EBRAINS cloud services in particular, helping users to reproduce several TVB publications. Tutorials consist of short video lectures, scripting tutorials with Jupyter notebooks and code. TVB Made Easy is a series of short lectures that introduce working with TVB in a clinical context, e.g., predicting recovery after stroke and simulating epilepsy patient brains. Brief lectures describe methods, results and ways to replicate the principal ideas of the articles with TVB.

Data protection in the TVB on EBRAINS cloud

Biomedical research is facing challenges because many methods lack technical infrastructure to protect the privacy of personal data. Problematically, biomedical data cannot be easily anonymized or pseudonymized such that all potentially identifiable information are removed, and potential re-identification is excluded (Byrge and Kennedy, 2018; Gymrek et al., 2013; Rocher et al., 2019). To use cloud services, personal data has to be transmitted over the internet and other open networks and is stored and processed on shared supercomputers, which poses the risk for unauthorised access of the data. The European Union's General Data Protection Regulation (GDPR) and similar international and national laws impose restrictions on the processing of personal data including the storage and sharing of data. From the outset of the project, at the development stages, Article 25 GDPR ("data protection by design and by default") requires partners to implement the appropriate technical and organizational measures, to ensure adherence with the principles of data protection at set out in Article 5 GDPR and more generally a commitment to GDPR compliant data processing to safeguard the rights and freedoms of the data subjects as set out in the GDPR. A principle means of ensuring GDPR compliant data processing is the implementation of appropriate technical and organizational measures to ensure a level of security appropriate to the risk of the processing as set out in Article 32 GDPR. An assessment of the most suitable security measures must consider the context and purposes of processing in relation to the risk posed to the rights and freedoms of the data subject. Having made this assessment, the appropriate security measures are realized for TVB-on-EBRAINS by implementing access control, encryption and sandboxing (Figure 2).

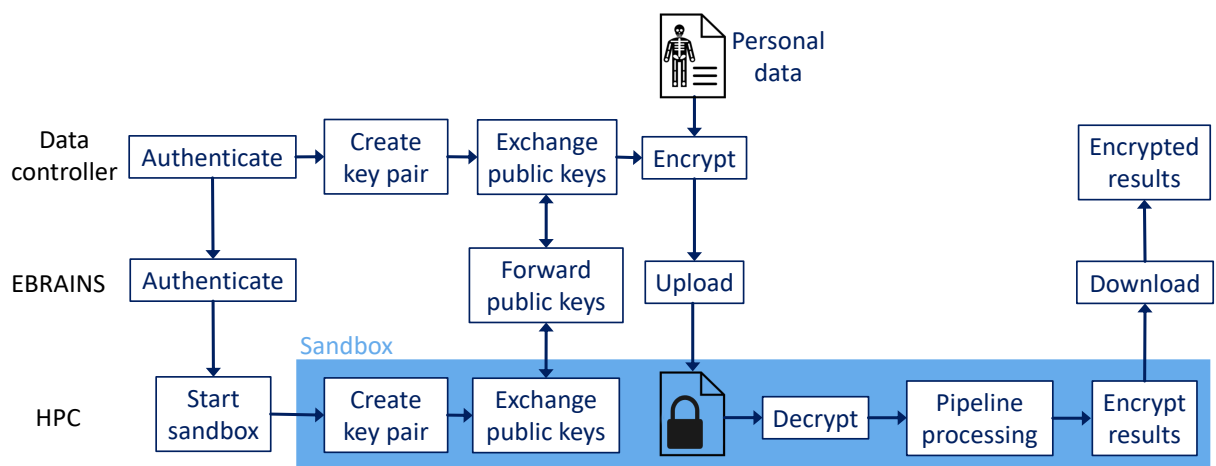

Figure 2. Securing personal data processing workflows in shared environments. Personal input data is encrypted with public key cryptography on the data controller's computer before upload to the cloud. The key pair for upload is generated within a sandboxed process at the final processing site and the private key never leaves the sandbox. All processing is performed in the sandbox and personal data is never written outside the sandbox in unencrypted form. A public key generated by the data controller is used for returning encrypted results.

EBRAINS access control uses passwords and cryptographic keys to prevent unauthorized access, to provide secure delegated access for connecting different cloud services, as well as to provide single sign-on. Keycloak is used for identity and access management (IAM), i.e., user registration, management and permission control. The OpenID Connect protocol is used for authentication--confirming the identity of a user--based on OAuth 2.0 specifications for delegating and conveying authorization decisions. OAuth 2.0 is a widely adopted standard for access delegation and user authorization flows without the need for sharing credentials, which is realized by issuing cryptographic tokens that provide ongoing access to protected resources on behalf of a user. With this framework EBRAINS applications can limit the scope of services accessible by a user.

Encryption is a fundamental tool for data privacy, ensuring that data becomes unintelligible without decryption key. Therefore, data is better protected if it is encrypted at all times with the only exception being during the time of the processing, where it is only decrypted for the parties involved; additionally, during the processing it may exist in unencrypted form only within isolated memory and file-system locations that are invisible from the host (sandboxes). TVB-on-EBRAINS workflows (Figure 2) require that personal data must already be encrypted on the data controller's computer before being uploaded to the cloud. Public key cryptography is used in a way that the secret key for data upload is created ad-hoc at the beginning of the workflow and remains the entire time in a sandboxed process at the final processing site. Processing results are never written out in unencrypted form, except in temporary filesystem trees that are invisible from the host and automatically cleaned up when the last process exits. For upload to the TVB Web GUI personal data is first encrypted with a public key and immediately after arrival in the cloud decrypted and again encrypted with a freshly created secret key to increase protection in case that the key for data upload is leaked. The data stays encrypted until an authorised user opens the respective project in its private TVB Web-GUI space. Each project is encrypted with a different key and the public key for initial encryption is regularly changed. For resource-intensive simulations the encrypted data is forwarded to supercomputing systems and only decrypted after the associated job on the compute nodes was started. Furthermore, the decryption key is sent from the TVB Web-GUI instance directly to the running backend compute process on the supercomputer using an authenticated security token; the key is only kept in memory, and never written to the file system. See Supplementary Note: Data protection in the TVB on EBRAINS cloud for more information.

Shared responsibility & compliance

In order to use TVB cloud services a user must agree to terms that clarify their personal responsibility regarding the compliance with the GDPR (ebrains.eu/terms). These terms transparently clarify the nature of security precautions, contact persons, personal responsibilities of the user, monitoring, logging and passing of information to third parties. A detailed risk assessment and technical documentation are currently in preparation to clarify inherent risks of open networks and shared computing systems and the taken countermeasures. Security measures were put in place to prevent that users are able to access another user's data if not explicitly shared.

The system is designed such that the user is the only person that is in control over the data while in the cloud. From a GDPR point of view they are therefore considered to be data controllers (Article 26 GDPR), which is the person determining the (essential) means and purposes of the processing, Hence, the parties using the cloud do not cede their legal responsibilities for data protection—on the contrary they remain unreservedly with the user as the data controller, because the technical and organisational mechanisms, that were put in place to prevent unauthorized access, guarantee their sole and independent determination on the purposes and means of the data processing operation.

Discussion

TVB cloud services were developed to lower the barriers to brain simulation and connectome analysis workflows and to enhance their reproducibility. All codes are open source and available for download from GitHub. Software is packaged in platform-independent container images that can be directly used without the need to install dependencies. Most software and data components were peer-reviewed, and results published in academic journals (Table 2). To ensure long-term accessibility, software libraries from third-party developers were forked and archived. Codes and images are versioned for reproducibility; legacy codes and images remain available in their repositories. By reporting the application name and version in a publication, it becomes possible to exactly replicate the used workflow, which counteracts variability between software versions and lack of reporting. Comprehensive documentation in the form of manuals, tutorials, lectures, Jupyter notebooks, demo data, workshops, videos, use cases, mailing lists and support contacts provide efficient and didactic dissemination of knowledge and support. EBRAINS core services enable to map and organize projects into a persistent and replicable structure at a central and secure place, which makes it easier to pick up complex projects at a later time. The flexibility of the platform and its focus on community-driven research enable rapid adoption of advances in brain simulation and connectomics, as well as correction of errors. Technical and organisational security mechanisms are designed to provide highest data protection standards, while at the same time providing flexibility to enable state-of-the-art research. To keep the high quality of the cloud services, ongoing and future efforts are directed towards the continuous integration of improved community standards and best practices. The TVB on EBRAINS ecosystem can be transferred to other Cloud environments within the European Open Science Cloud or beyond. Thus, it serves as a reference architecture for secure processing and simulation of neuroscience data in the cloud (Figure 1 and Supplementary Discussion).

Methods

TVB Image Processing Pipeline

The EBRAINS TVB Image Processing Pipeline combines three BIDS Apps (see Supplementary Note: BIDS Apps) to yield an updated version of our brain network model construction workflow (Schirner et al., 2015). BIDS Apps are brain imaging software packages that understand BIDS datasets and that are deployed as portable container images (Gorgolewski et al., 2017, 2016):

- thevirtualbrain/tvb-pipeline-sc for diffusion MRI tractography,

- thevirtualbrain/tvb-pipeline-fmriprep for fMRI preprocessing, and
- thevirtualbrain/tvb-pipeline-converter for data privacy and conversion of the results of the other two containers into TVB and BIDS formats.

All three container images are provided on TVB's main Docker Hub Repository hub.docker.com/r/thevirtualbrain. The code for the container tvb-pipeline-sc was cloned and modified from the BIDS App MRtrix3_connectome (github.com/BIDS-Apps/MRtrix3_connectome) and tvb-pipeline-fmriprep was cloned from the neuroimaging software fmriprep (github.com/poldracklab/fmriprep), which is why we recommend to acknowledge MRtrix3, MRtrix3_connectome (Smith and Connelly, 2019; Tournier et al., 2019) and fmriprep (Esteban et al., 2019) when using these workflows. Containerization makes it easier to deploy these workflows on different architectures, as they rely on a number of dependencies.

The overall workflow is coordinated by a central program that orchestrates the execution of the three containers on a supercomputer and that ensures that personal data is encrypted at all times, except for the duration of the processing and then only in the main memory of a sandboxed process (Figure 2). All processing of personal data on the supercomputer happens inside the sandbox and intermediary data is only written out into a temporary file system that is invisible from the host. After a data controller authenticated with the EBRAINS platform and initiated a processing workflow, EBRAINS authenticates with the supercomputer and starts a sandboxed process that is isolated from the host using Bubblewrap (github.com/containers/bubblewrap), which is a sandboxing tool based on Linux user namespaces that allow to give unprivileged users container features. Bubblewrap creates a new mount namespace where the root is on a temporary filesystem that is invisible from the host and that will be automatically cleaned up when the last process exits. Shortly before the upload of the encrypted data a pair of public and private keys is generated, and the public key is forwarded to the data controller to encrypt the password for the encrypted input data. The data themselves is encrypted on the data controller's computer with AES-256 encryption using pyAesCrypt (pypi.org/project/pyAesCrypt). After the encrypted data and password were uploaded to the supercomputer, the password and data are decrypted into the sandboxed temporary file system. The secret private key for decryption is only held in the main memory of the sandboxed process and never written out. During the entire processing intermediate results are only written into the temporary filesystem that is invisible from the host. The outputs of the workflow are encrypted with a public key that was generated on the data controller's computer and all input and intermediate data of the workflow are deleted, the sandbox stopped, and the encrypted results returned to the data controller.

Processing of functional or diffusion-weighted MRI requires at least one high-resolution T1-weighted structural MRI for parcellating the brain. If the input data was not corrected for susceptibility distortions, then additionally phase-reversed images of the B0 field in dwMRI data and fieldmaps for fMRI are required. Please see the Supplementary Note: Exemplary input data for more information on input data sets.

thevirtualbrain/tvb-pipeline-sc

Diffusion modelling and tractography are carried out by MRtrix3, which is a powerful and actively developed tractography toolbox (Tournier et al., 2019). Notably, it includes methods to increase the accuracy of tractography and reducing false positives by removing tracks that are anatomically implausible, called Anatomically-Constrained Tractography (Smith et al., 2012). In order to estimate region-to-region coupling strengths, the MRtrix3 programs SIFT, respectively SIFT2, re-weight each track in the reconstructed tractogram such that streamline densities become proportional to the cross-sectional area connecting each pair of brain regions (Smith et al., 2015).

The workflow is controlled by a Python script that can be modified according to the research question. In the following we describe the default setup of the workflow as it is currently implemented. As input, the workflow requires T1-weighted and diffusion-weighted MRI data, as well as at least one phase-reversed dwMRI of the B0 field for susceptibility distortion correction. Outputs are whole-brain tractograms and SC. In between, the following processing steps are performed. First, dwMRI is denoised by removing noise-only principal components (Veraart et al., 2016) using *dwidenoise*. Next, Gibbs ringing artifacts (Kellner et al., 2016) are removed using *mrdegibbs*, and distortions are corrected with *dwipreproc*: eddy current-induced distortion correction and motion correction is performed using FSL *eddy*, and (optionally) susceptibility-induced distortion correction is performed using FSL *topup*. In the next step, B1 bias field inhomogeneities are corrected using *dwibiascorrect* and a brain mask for DWI is created using *dwi2mask* and *maskfilter*. After computing fractional anisotropy maps using *dwi2tensor*, *dwi2response* is used to estimate the response functions for spherical deconvolution. To obtain fiber orientation distributions, spherical deconvolution is performed with *dwi2fod*. For inter-modal registration, the workflow extracts the brain with *ROBEX* and performs a bias field correction on the T1 image in its original space using ANTS *N4BiasFieldCorrection*. Contrast-matched images for inter-modal registration between DWIs and T1 are generated using *mrhistmatch* and the T1w MRI is registered to DWI using *mrregister*. The MRtrix program *5ttgen* in conjunction with FSL or FreeSurfer is then used to segment tissues into cortical grey matter, sub-cortical grey matter, white matter and cerebrospinal fluid. Next, cortical gray matter parcellations are obtained. For the atlases 'desikan', 'destrieux', and 'hcpmmp1', the parcellation information is taken from FreeSurfer's *recon-all* output. For the atlases 'aal', 'aal2', 'craddock200', 'craddock400', and 'perry512', a non-linear registration to the MNI152_T1_2mm template is performed with optionally ANTS or FSL, and the registration result is used to transform the atlas information into individual subject space. Based on this nonlinear registration any atlas defined on the MNI152 template can be used to parcellate the subject's brain, e.g., cytoarchitectonic information from Julich-Brain. Next, the central step in this workflow is performed: whole-brain fibre-tractography using *tckgen* with ACT to control for anatomical plausibility of constructed tracks using information from the tissue-segmented image (Smith et al., 2012). As additional plausibility criteria, tracks are truncated and re-tracked if a poor structural termination is encountered, respectively cropped when they hit the gray-matter-white-matter interface and when they exceed a length of 250 mm. Seed points are determined dynamically using the SIFT model (Smith et al., 2015). If the number of tracks to be generated is not manually specified, it is set to 500*N*(N-1), where N is the number of brain regions in the parcellation. After a whole-brain tractogram has been generated, SIFT2 (Smith et al., 2015) is used next to filter tractography results to determine streamline weights. SIFT2 optimises per-streamline cross-section multipliers to match a whole-brain tractogram to fixel-wise fibre densities. As last step of the MRtrix workflow, *tck2connectome* is used to aggregate the whole brain tractogram into a region-by-region connectome matrix using the specified region parcellation. In addition to outputting the sum of streamline weights for each region pair, also the mean streamline lengths between each region is output. Optionally, the workflow also outputs track density images, which are useful for visualising and evaluating the results of tractography.

thevirtualbrain/tvb-pipeline-fmriprep

The fMRI workflow relies on fmriprep, which is a preprocessing pipeline for fMRI data that is relatively robust to variation in scan acquisition protocols, sequence parameters, or presence of fieldmaps for artifact correction (Esteban et al., 2020, 2019). A characteristic feature of fmriprep is its "glass box" philosophy, according to which reports for visual verification are produced after important processing steps. This is an important feature as many steps of MRI preprocessing pipelines are susceptible to errors and verification is therefore generally recommended after major steps. The workflow combines the following software: FreeSurfer, FSL, AFNI, ANTS, BIDS validator and ICA-AROMA. Auxiliary tools are: Python (miniconda), pandoc, SVGO, neurodebian and git. As input the workflow requires data to be in BIDS format, and it must include at least one T1w structural image and one fMRI series. Output contains preprocessed fMRI data, as well as noise components estimated with independent component analysis, which are used to remove associated variance from the fMRI data during the tvb-pipeline-converter workflow. The following processing steps are performed: T1w volumes are corrected for intensity non-uniformity using *N4BiasFieldCorrection* and skull-stripped using *antsBrainExtraction.sh*. Brain surfaces are reconstructed using FreeSurfer's *recon-all*, and the temporary brain mask is refined to reconcile ANTs-derived and FreeSurfer-derived segmentations of the cortical gray-matter of Mindboggle. Spatial normalization to the ICBM 152 Nonlinear Asymmetrical template version 2009c is performed through nonlinear registration with the *antsRegistration* tool, using brain-extracted versions of both T1w volume and template. Brain tissue segmentation of cerebrospinal fluid (CSF), white-matter (WM) and gray-matter (GM) is performed on the brain-extracted T1w using FSL *FAST*. Functional data is slice time corrected using *3dTshift* from AFNI and motion corrected using *mcflirt* from FSL. Distortion correction is performed using an implementation of the TOPUP technique using *3dQwarp*. This is followed by co-registration to the corresponding T1w using boundary-based registration with nine degrees of freedom, using *bbregister* from FreeSurfer. Motion correcting transformations, field distortion correcting warp, BOLD-to-T1w transformation and T1w-to-template (MNI) warp are concatenated and applied in a single step using *antsApplyTransforms* using Lanczos interpolation. Physiological noise regressors are extracted applying CompCor. Principal components are estimated for the two CompCor variants: temporal (tCompCor) and anatomical (aCompCor). A mask to exclude signal with cortical origin is obtained by eroding the brain mask, ensuring it only contained subcortical structures. Six tCompCor components are then calculated including only the top 5% variable voxels within that subcortical mask. For aCompCor, six components are calculated within the intersection of the subcortical mask and the union of CSF and WM masks calculated in T1w space, after their projection to the native space of each functional run. Frame-wise displacement is calculated for each functional run using the implementation of Nipype. ICA-based Automatic Removal Of Motion Artifacts (*AROMA*) was used to generate aggressive noise regressors as well as to create a variant of data that is non-aggressively denoised (see section thevirtualbrain/tvb-pipeline-converter for details on aggressive vs. non-aggressive cleaning). For details, please refer to fmriprep's documentation (https://fmriprep.readthedocs.io/).

thevirtualbrain/tvb-pipeline-converter

The third container of the TVB image processing pipeline takes the outputs of the first two containers as input and outputs TVB-ready BNM connectomes in BIDS format (Supplementary Table 1) and in native TVB-ready format (Supplementary Table 2). Outputs include SC, FC, EEG/MEG (magnetoencephalography) projection matrices, and brain surface triangulations. As first step, the workflow performs what fmriprep documentation calls "non-aggressive" cleaning of motion artefacts. Following its glass box philosophy, fmriprep involves the researcher in critical decisions about the denoising strategy, as it can have considerable effects on the ensuing analyses. To this end, fmriprep provides both aggressively and non-aggressively denoised fMRI. The output of the fmriprep workflow already contains ICA-AROMA denoised 4D NIFTI files mapped to MNI space. Here, the converter container performs non-aggressive denoising following the approach described in Pruim et al. (2015) using the independent components (ICs) together with their classification obtained in the previous step. ICA-AROMA classifies ICs into movement-related and movement-unrelated ICs. It does not require manual training of the classifier--the classifier is already trained using four theoretically motivated spatial and temporal features: high-frequency content, correlation with realignment parameters, edge fraction, and CSF fraction of each IC (Pruim et al., 2015). With aggressive denoising the entire temporal waveform of movement-related ICs is regressed from the data, analogous to the commonly done regression of nuisance parameter time courses (like mean global signal, mean tissue class signal, frame-wise displacement, motion parameters) from the data. Critically, all variance associated with such "nuisance" time courses is removed, including shared variance with actual brain signals. This is problematic, because global fMRI time courses often share a lot of variance with neural activity: even if two signals originate from different locations in the brain, they may still be highly correlated. Therefore, regressing out global signals likely removes signals of interest (Liu et al., 2017; Murphy and Fox, 2017). As an alternative, the non-aggressive approach based on ICA-AROMA is more conservative by first performing a regression on the full set of IC time series, including both signal and noise ICs. Then, to clean the data, only the motion-related regressors are subtracted from the data, which specifically removes variance associated with motion-related-ICs that is not shared by the remaining ICs. Problematically, with this non-aggressive approach motion-related dynamics that were not identified as such are specifically retained in the data. Conversely, as mentioned, the drawback of aggressive denoising is that it may remove shared variance between signal and noise. To clean fMRI data non-aggressively the workflow uses *fsl_regfilt* with the provided AROMA ICs and a list of (motion-related) ICs to reject. After this step, further nuisance regression may be performed (if it fits the study design) as well as removal of linear trends and high-pass filtering. It is, however, important that during such a second regression step variables correlated with motion are not used as regressors, as this may re-introduce motion-related waveform components into the time series.

After denoising, the converter computes region-average fMRI time courses by resampling the brain parcellation to fMRI resolution and averaging over all voxel time series in each region. The converter also outputs cortical surfaces, which are used for detailed surface simulations and for EEG or MEG source modelling, e.g. in order to extract region-average EEG or MEG source activity, which can be used to constrain BNM dynamics (Schirner et al., 2018). The converter merges the left and right hemisphere cortical surface triangulations reconstructed by FreeSurfer and generates a mapping between each vertex and the large-scale regions of the brain atlas. Each vertex of the cortical white matter and pial surface triangulations are associated with the region-label of the region at that location. This step is carried out with HCP Connectome Workbench for atlases that are defined in volumes (e.g., "aal", "aal2" "craddock200", "craddock400", "perry512", atlas names as used in MRtrix workflow source codes) and outputs a GIFTI label file for each hemisphere. For atlases defined on surfaces (e.g., "desikan", "destrieux", "hcpmmp1"), the 'annot' files from FreeSurfer are used to obtain the region mapping. The following atlases are currently natively supported, for other atlases the workflow needs to be adapted: "aal", "aal2", "craddock200", "craddock400", "desikan", "destrieux", "hcpmmp1", "perry512".

Next, the MNE toolbox (Gramfort et al., 2013) is used to compute forward and inverse models for electromagnetic source imaging. First, surfaces are decimated to 30,000 triangles and the region-mapping is obtained by nearest neighbor interpolation in the original high-resolution surface. For the head model, BEM surfaces using the FreeSurfer watershed algorithm are constructed. To make the workflow automatic, standard EEG montage locations are projected onto the individual head surfaces. The outputs of this step are a projection or lead-field matrix (LFM), EEG sensor location coordinates, and a mapping between surface vertices and large-scale regions. The LFM has the dimensions $M \times N$, with $M$ being the number of vertices on the cortical surface and $N$ the number of EEG sensors. Next, surfaces are exported: the downsampled pial surface from FreeSurfer (used to define source space) as well as the BEM surfaces of the inside and the outside of the skull and the scalp. Finally, the converter outputs the SC weights and distances matrices as well as region centroids, average orientations (average over all vertex-normals), surface areas, and two files that indicate to which hemisphere each region belongs and whether it is a cortical or a subcortical region. All output files are plaintext ASCII files that are zipped for uploading to TVB. As last step, the converter generates metadata for the BIDS output (see Supplementary Note: Metadata annotations). Supplementary Table 1 summarizes the folder and filename structure in BIDS format and Supplementary Table 2 summarizes folder and filename structure for outputs in TVB format, which are all stored in the output file "TVB_output.zip" that can be imported into TVB using the "Upload Connectivity ZIP" functionality of TVB.

Multiscale Co-Simulation

Multiscale Co-Simulation is currently implemented in the form of two toolboxes: the TVB-Multiscale toolbox and the Parallel CoSimulation toolbox (Table 1). Both are based on common concepts and architectures, but they are independently developed to focus on different goals: TVB-Multiscale focusses on rapid prototyping of scientific use cases while Parallel CoSimulation focusses on the optimization of co-simulation performance.

In order to couple the small scale (NEST) with the large scale (TVB), TVB NMM equations are coupled with NEST single neuron equations. The coupling can be unidirectional, i.e., one scale provides input to the other scale, or bidirectional, i.e., both scales provide input to one another. The bidirectional case enables to "substitute" large-scale NMMs by small-scale spiking networks to simulate one or more specific populations of a BNM on a finer level. Large-scale inputs to the substituted populations will then be forwarded to the small-scale network, while small-scale activity will be averaged, or transformed with a custom-defined transformation function, and forwarded as input to the connected large-scale nodes. For example, such a transformation function may convert moment-to-moment firing rates of a large-scale NMM into spike trains by sampling from Poisson distributions centered at these firing rates, which can then be used to drive spiking network models. Vice versa, spike trains of a population of neurons may be used to compute instantaneous population firing rates, which are then used to drive large-scale NMMs. Like in the case of large-scale-only BNMs, it is possible to subsequently input simulated neural activity into forward models to simulate typical neuroimaging signals like fMRI or EEG. The TVB to NEST interface is based on the creation of TVB "proxy" nodes within the spiking network model. Proxy nodes are NEST stimulation devices that are used to inject spikes or currents into NEST neurons in order to simulate inputs from the large scale to the small scale at a particular brain region. It is possible to generate currents or spike trains with a desired first order (mean firing rate) or second order statistics (correlations) such that the statistics of the produced spikes or currents follow the corresponding statistics of the large-scale population activity. Proxy nodes can be coupled to NEST spiking networks with user defined connection weights and delays, in order to simulate the large-scale features of coupling on the small scale. To compute inputs from NEST to TVB, NEST recording devices are used to aggregate the activity of spiking neuron populations. Multiscale Co-Simulation gives the user flexibility for custom configuration of co-simulations regarding the network structures, the state variables used for coupling, the transformation functions and the devices for computing and applying inputs, the allocation of computational resources, and the storage of results.

The Parallel CoSimulation toolbox is currently under development to optimize the performance of co-simulation as well as for integration with the TVB-Multiscale toolbox. The idea for optimization is to reduce the number of costly communication operations between NEST and TVB, because inputs often do not need to be exchanged in every single time step of the model integration. Rather, it is often the case that the model does not require instantaneous interactions, depending on the axonal transmission delays in the network. The Parallel CoSimulation toolbox therefore employs a strategy where each simulator integrates their respective model equations independently and in parallel for a number of time steps, until again an exchange of inputs becomes unavoidable. The toolbox uses MPI (Message Passing Interface) for communication and is implemented in a modular fashion: independent modules for simulation and for input transformation enable scaling over multiple compute nodes and to allocate different computational resources to each module. The communication between simulators is implemented using Message Passing Interface (MPI), and synchronization is taken care by the mediating transformation modules, in which the receiving data are also adapted to the appropriate inputs' format of each simulator. The simulators and the transformers are encapsulated in independent modules allowing for flexibility regarding the components of any particular instance of co-simulation, as well as for scaling co-simulation to multiple supercomputer nodes. A benchmark of the toolbox is provided in Supplementary Note: TVB Multiscale Co-Simulation benchmark.

High-Performance implementations of TVB

For TVB-HPC the domain-specific language RateML was developed, which allows to specify custom NMMs without requiring knowledge about how to optimally implement such models. Python code for CPU and CUDA code for GPU is automatically produced from RateML model specifications. The resulting Python code uses Numba vectorization (Lam et al., 2015) to integrate the NMMs. The resulting CUDA code uses the highly parallel architecture of modern GPUs by spawning one thread for every simulated parameter set, allowing to simulate multiple models in parallel. RateML is based on the domain-independent language 'LEMS' (Vella et al., 2014), which supports the declarative description of model components in a concise XML representation. The PyLEMS expression parser is used to check and parse mathematical expressions (Vella et al., 2014). Exemplary RateML ports of the TVB NMMs Epileptor, Kuramoto, 2D Oscillator, Montbrio and ReducedWongWang are provided with the main TVB software package.

Code with model equations is mapped to *generalized universal function* (gufuncs) using Numba's *guvectorize* decorator, which compiles a pure Python function directly into machine code that can be used to operate on NumPy arrays. An example of a Numba generated *guvectorize* function is displayed in Listing 1. In this example the *gufunc _numba_dfun_Epileptor* accepts two *n* and *m* sized float64 arrays and 19 scalars as input and returns a *n* sized float64 array. The benefit of writing NumPy gufuncs with Numba's decorators, is that it automatically uses parallel operations such as reduction, accumulation and broadcasting to efficiently implement an algorithm. Please see Supplementary Methods: TVB-HPC for further notes on the implementation and a benchmark.

```
@guvectorize([(float64[:],
float64[:],    (float64    *    19),
float64[:])],
 '(n),(m)' + ',()'*19 + '->(n)',
nopython=True)
def    _numba_dfun_EpileptorT(vw,
coupling, a, b, c, d, r, s, x0, Iext,
slope,
  Iext2, tau, aa, bb, Kvf, Kf, Ks, tt,
modification, local_coupling, dx):

  c_pop1 = coupling[0]
  c_pop2 = coupling[1]
  c_pop3 = coupling[2]
  c_pop4 = coupling[3]
  ... # calculate derivatives
```

```
    return dx
```
Listing 1. Example of a gufunc header for the epileptor model, where vw holds the input and dx the computed output derivatives.

Fast_TVB was developed in C and makes extensive use of several optimization techniques to increase the speed of BNM simulation. It uses Single Instruction, Multiple Data instructions where the same operation is concurrently applied to multiple values contained in one large register. In the central integration loop no function calls are made to avoid possible overhead and instead all necessary code is either directly written into the loop or inlined. Inner loops are unrolled, jammed and bound to avoid "end of loop" tests and branch penalties. Intermediary results are re-used in other parts of the algorithm and not computed again. For example, some terms and expressions occur multiple times in the equations or they resolve to a constant and therefore only need to be computed once and not repeatedly in each time step. To reduce memory-related performance bottlenecks and cache failures, data to compute coupling inputs is organized in an efficient ring-buffer that stores related data in close locality. SC is stored in a sparse layout that scales linearly with the number of connections (and not quadratically as when stored as array), which enables to hold large BNMs efficiently in memory. The computation of coupling inputs also requires that the program steps through distant memory locations (in order to fetch time-delayed state variables), which is why pointer tables that directly link to the required addresses are used instead of array indexing tables, which saves one dereferencing operation for each memory access. Input and output operations happen only before or after the main simulation to avoid waiting time for devices. Further information and a benchmark are provided in Supplementary Note: Fast_TVB.

Bayesian Virtual Epileptic Patient

Bayesian inference is used to compute posterior probability distributions for parameters of TVB's Epileptor NMM (Hashemi et al., 2020; Jirsa et al., 2017, 2014; Proix et al., 2017). The Epileptor was developed on the basis of a taxonomy analysis of seizure dynamics that has shown that a system of five linked state variables is sufficient to describe onset, time course and offset of ictal-like discharges and their recurrence (Jirsa et al., 2014). A *detailed analysis* over 2000 focal-onset seizures from multiple centers has shown that the Epileptor model is able to realistically simulate the most dominant dynamics in seizure-like events (Saggio et al., 2020). The target of inference is to obtain posterior distributions of the excitability parameter for every brain region in order to assemble patient-specific epileptogenicity maps; furthermore, initial conditions, global coupling, and noise parameters are fitted (in total 3N+3 parameters for N brain regions). The excitability parameter controls whether the Epileptor shows epileptogenic behavior or not and the estimated values are used to classify brain regions into three categories: epileptogenic zones (EZ), which can autonomously trigger seizures; propagation zones (PZ), which do not trigger seizures autonomously but may be recruited during the seizure evolution; and healthy zones (HZ), where no seizures occur.

As exact posterior inference in such high-dimensional models is intractable, approximate methods are used, e.g., Markov Chain Monte Carlo (MCMC), which enables to generate correlated samples that converge to a (potentially complex) target distribution. Gradient-based MCMC algorithms like Hamilton Monte Carlo provide efficient convergence to high-dimensional target distributions, but their performance is highly sensitive to hyperparameters, which often need to be re-tuned to arrive at the desired target distribution, which is solved in BVEP by using No-U-Turn Samplers for adaptive self-tuning of hyperparameters (Hoffman and Gelman, 2014). Alternatively, Automatic Differentiation Variational Inference (ADVI) is used to approximate the posterior by first positing a family of densities and then finding a member of that family that is closest to the target distribution as measured by Kullback-Leibler divergence (Kucukelbir et al., 2017). BVEP allows to integrate clinical information as priors, such as a patient's SC, or lesions detected in MRI. Priors for the excitability parameter are either uninformative (e.g., a uniform distribution over the entire range for EZ, PZ and HZ), or they can express clinical hypotheses (e.g., by extending a distribution only over one category) or for including observations (e.g., when the region was recorded to show seizure activity, the range for HZ can be excluded). The fitting result is evaluated using information criteria and approximate leave-one-out cross-validation metrics to assess the model's ability in predicting data it was not trained for. Model evidence is used to compare different clinical hypotheses and to further inform physicians before therapeutic interventions. BVEP is implemented in the open-source probabilistic programming language tools Stan (mc-stan.org) and PyMC3 (docs.pymc.io), which use automatic differentiation to compute gradients of specified model density functions for NUTS and ADVI. The workflow was tested with ground-truth synthetic data for two patients with symptomatic and asymptomatic seizures where it was possible to accurately and efficiently infer the spatial map of epileptogenicity for all brain regions (Hashemi et al., 2020). See Supplementary Note: Bayesian Virtual Epileptic Patient for more information.

TVB Mouse Brains

For TVMB (Melozzi et al., 2019, 2017) tracer-based mouse SC at different resolutions was constructed from the Allen Mouse Brain Connectivity Atlas (Oh et al., 2014) with a maximum resolution of 50 µm, yielding 540 brain regions. The "Mouse Stroke" workflow associates structural alterations related to stroke with corresponding changes in FC and population synchronization (Allegra Mascaro et al., 2020). To model stroke and subsequent rehabilitation, a pre-defined fraction of incoming connections is removed, e.g., 0 % corresponds to no stroke, while 100 % would correspond to a devastating stroke that destroyed all incoming connections to the region. Similarly, recovery from stroke is modelled by compensatory rewiring (Nudo, 2013) where the connection strengths of the non-damaged connections are increased relative to healthy connectivity. Different stroke connectomes are produced in this manner and the resulting FC is then fitted and compared with empirical data. By testing different re-wiring scenarios, the relative importance of compensatory rewiring in different connections for re-establishing healthy FC is studied. For these BNMs the Kuramoto model (Kuramoto, 1984) was used, which is a phenomenological model for emergent group dynamics of weakly coupled oscillators that we used previously to link SC with oscillatory dynamics in different modalities (Petkoski et al., 2018; Petkoski and Jirsa, 2019).

## TVB-ready data

*Ethics statement.* The data analyzed here have been reported in previous studies (Aerts et al., 2020, 2018). Both studies were approved by the Ethics Committee at Ghent University Hospital. All participants received detailed study information and gave written informed consent before study enrolment.

This dataset contains MRI derivatives from patients who were diagnosed with either a glioma, developing from glial cells, or a meningioma, developing in the meninges, as well as healthy controls. Patients were recruited from Ghent University Hospital (Belgium) between May 2015 and October 2017 on the day before each patient's tumor surgery. Patients were eligible if they (1) were at least 18 years old, (2) had a supratentorial meningioma (WHO grade I or II) or glioma (WHO grade II or III) brain tumor, (3) were able to complete neuropsychological testing, and (4) were medically approved to undergo MRI investigation. Partners were also asked to participate in the study to constitute a group of control subjects that suffer from emotional distress comparable to that of the patients. Data from 11 glioma patients (mean age 47.5 y, SD = 11.3; 4 females), 14 meningioma patients (mean age 60.4 y, SD = 12.3; 11 females), and 11 healthy partners (mean age 58.6 y, SD = 10.3; 4 females) was collected. From all participants, three types of MRI scans were obtained using a Siemens 3T Magnetom Trio MRI scanner: T1-MPRAGE anatomic images, resting-state functional echo-planar imaging data, a multishell high angular resolution diffusion-weighted MRI scan, and two DWI b = 0 s/mm2 images were collected with reversed phase-encoding blips for the purpose of correcting susceptibility-induced distortions. Further information on the dataset, preprocessing and analysis can be found in the original journal articles using this dataset (Aerts et al., 2020, 2018) and in the dataset publication on EBRAINS KnowledgeGraph (Aerts et al., 2019).

## Data availability

The datasets generated during and/or analysed during the current study are available in the EBRAINS KnowledgeGraph repository, search.kg.ebrains.eu. KnowledgeGraph is a DOI-minting repository where EBRAINS data and software services are indexed and/or archived. To share personal data among researchers subject to national or international data protection laws (e.g., the General Data Protection Regulation of the European Union) protected environments and workflows as well as data sharing agreement templates were created (Supplementary Note: Data protection in the TVB on EBRAINS cloud).

## Code availability

All software codes presented in this article have open-source licenses and can be freely downloaded from GitHub (Table 1). The EBRAINS database service KnowledgeGraph (search.kg.ebrains.eu) is a DOI-minting repository where all data and software services are indexed and/or archived. Source codes are deployed as cloud services that can be used on ebrains.eu and as standalone download versions that can be pulled as container images from Docker Hub (Table 1).

## Acknowledgments

We gratefully acknowledge the Swiss National Supercomputing Center CSCS for supporting this project by providing computing time through the Interactive Computing E-Infrastructure (ICEI) on the Supercomputer PizDaint of the Fenix Infrastructure (projects ich10, ich12) and the Gauss Centre for Supercomputing e.V. (www.gauss-centre.eu) for supporting this project by providing computing time through the John von Neumann Institute for Computing (NIC) on the GCS Supercomputer JUWELS at Jülich Supercomputing Centre (JSC). We acknowledge support by H2020 Research and Innovation Action grants Human Brain Project SGA2 785907, SGA3 945539, VirtualBrainCloud 826421 and ERC 683049; German Research Foundation CRC 1315, CRC 936, CRC-TRR 295, RI 2073/6-1, RI 2073/9-1 and RI 2073/10-2; Berlin Institute of Health & Foundation Charité, Johanna Quandt Excellence Initiative. Several computations have also been performed on the HPC for Research cluster of the Berlin Institute of Health. We acknowledge the use of Fenix Infrastructure resources, which are partially funded from the European Union's Horizon 2020 research and innovation programme through the ICEI project under the grant agreement No. 800858. German Research Foundation SFB 1436 (project ID 425899996); SFB 1315 (project ID 327654276); SFB 936 (project ID 178316478); SFB-TRR 295 (project ID 424778381); SPP Computational Connectomics RI 2073/6-1, RI 2073/10-2, RI 2073/9-1.

Author Contributions

Conceptualization, PR, VJ, JM, KA, JGB, TD, CCH, ARM, GD, MS; Data curation, DM, LS, RP, MS, LZ, OS; Formal analysis, MS, DM, LS, RP, PT; Funding acquisition, PR; Methodology, MS, DP, PT, MvdV, SD-P, AP, WK, OS, MW, LZ, JF, SP, LK, MH, J-FM, BS, TD, JM, PR; Project administration, PR; Resources, PR, VJ, JM, KA, JGB, TD, CMM, AU, MM, CCH, AF, J-FM, DM, DP, AN; Software, MS, LD, DP, PT, PP, BV, MvdV, SD-P, AP, WK, OS, MW, LZ, JF, SP, LK, MH, BS, TD, JM, GD; Supervision, PR, VJ, JM, KA, JGB, TD, CMM, MM, CCH, ARM, GD, BCS, SA, MC, EJ; Visualization, JP, CL, AB, PR; Writing – original draft, MS with contributions from all co-authors.

Competing interests

The authors declare no competing interests.